\documentclass[a4paper]{article}

\usepackage{INTERSPEECH_v2}

\usepackage{multirow}
\usepackage{subfig}

\title{Towards Lipreading Sentences with Active Appearance Models}
\name{George Sterpu, Naomi Harte}
\address{Sigmedia, ADAPT Centre, School of Engineering, Trinity College Dublin, Ireland}
\email{sterpug@tcd.ie, nharte@tcd.ie}

\begin{document}

\maketitle
\begin{abstract}

Automatic lipreading has major potential impact for speech recognition, supplementing and complementing the acoustic modality. Most attempts at lipreading have been performed on small vocabulary tasks, due to a shortfall of appropriate audio-visual datasets. In this work we use the publicly available TCD-TIMIT database,  designed for large vocabulary continuous audio-visual speech recognition. We compare the viseme recognition performance of the most widely used features for lipreading, Discrete Cosine Transform (DCT) and Active Appearance Models (AAM), in a traditional Hidden Markov Model (HMM) framework.  We also exploit recent advances in AAM fitting. We found the DCT to outperform AAM by more than 6\% for a viseme recognition task with 56 speakers. The overall accuracy of the DCT is quite low (32-34\%). We conclude that a fundamental rethink of the modelling of visual features may be needed for this task.

\end{abstract}

\section{Introduction}

Lipreading is the process of inferring someone's speech by analyzing the movement of their lips. 
Humans use lipreading to assist their auditory perception in tasks such as speaker localization, voice activity detection and ultimately speech recognition \cite{Sum87}. This skill allows a robust perception of speech in noisy acoustic environments, or when the hearing abilities have been partially or completely lost.

An open research problem in this area is finding the right representation of visual speech. As outlined by previous reviews \cite{review_pota, review_zhou}, most attempts demonstrate an improvement of the audio-visual fusion over the auditory-only modality, yet these results are generally valid for restricted tasks given the known limitations of the used datasets \cite{tcdtimit}. The main challenges come from speaker variation, pose variation and adequate exploitation of the temporal correlations \cite{review_zhou}, in addition to the context variation that causes co-articulation. Humans also rely heavily on their language skills when guessing difficult words or long sentences, so a proper integration of language, video and audio is required to reach human-level recognition performance.

\begin{figure}[t]
\centering
\subfloat[Holistic no-op]{
    \includegraphics[width=0.5\linewidth]{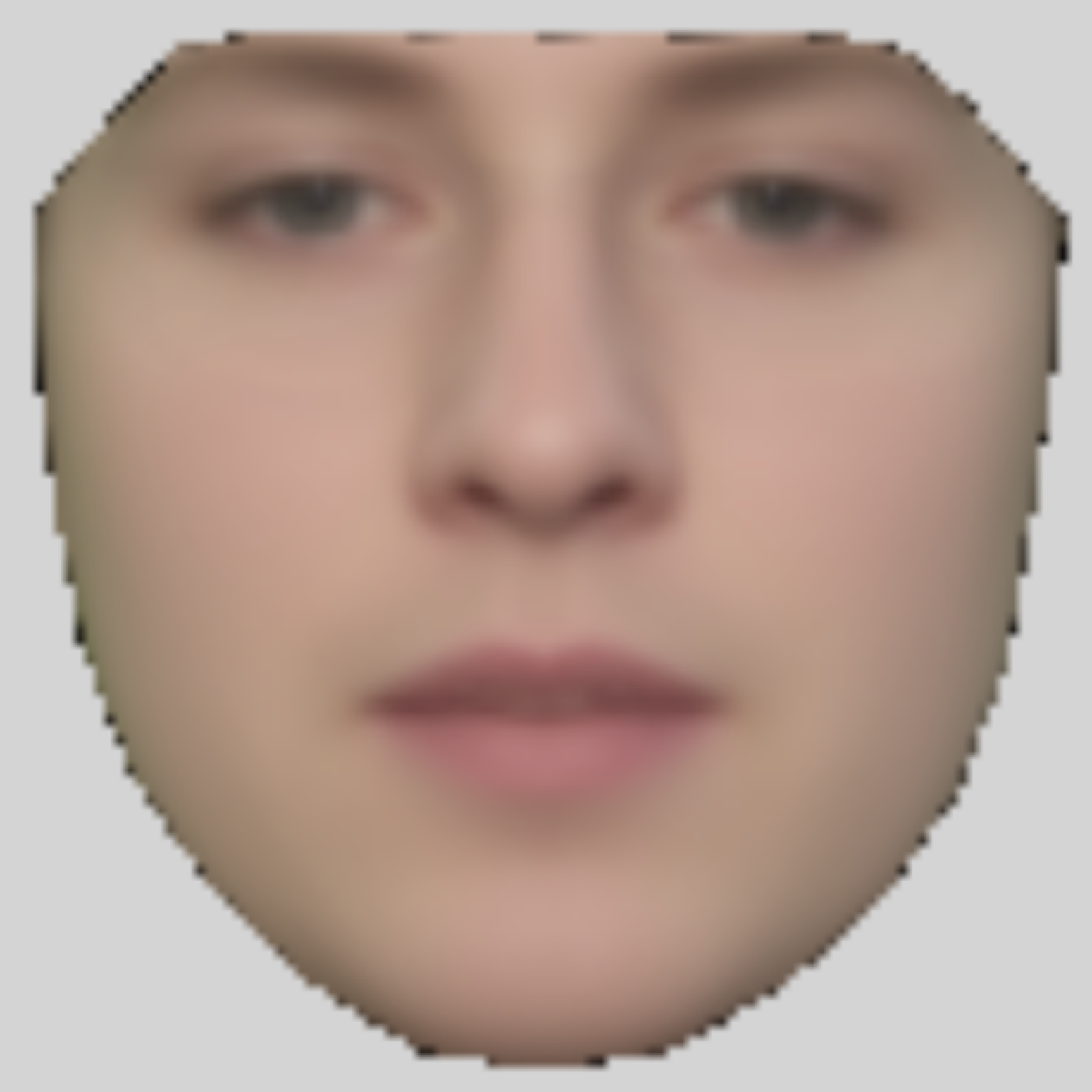}
}
\subfloat[Holistic SIFT]{
	\includegraphics[width=0.5\linewidth]{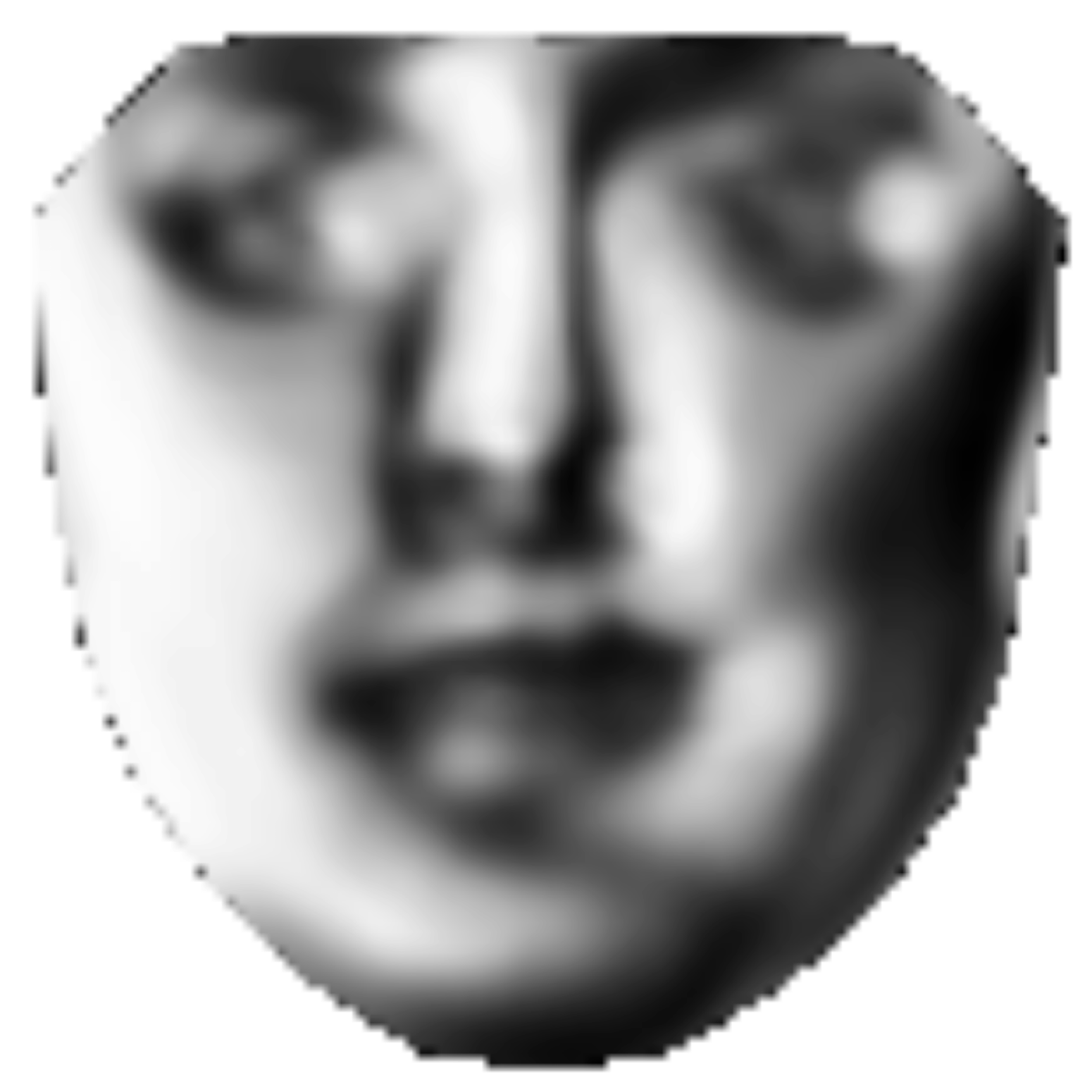}
}
\newline
\subfloat[Patch no-op]{
	\includegraphics[width=0.5\linewidth]{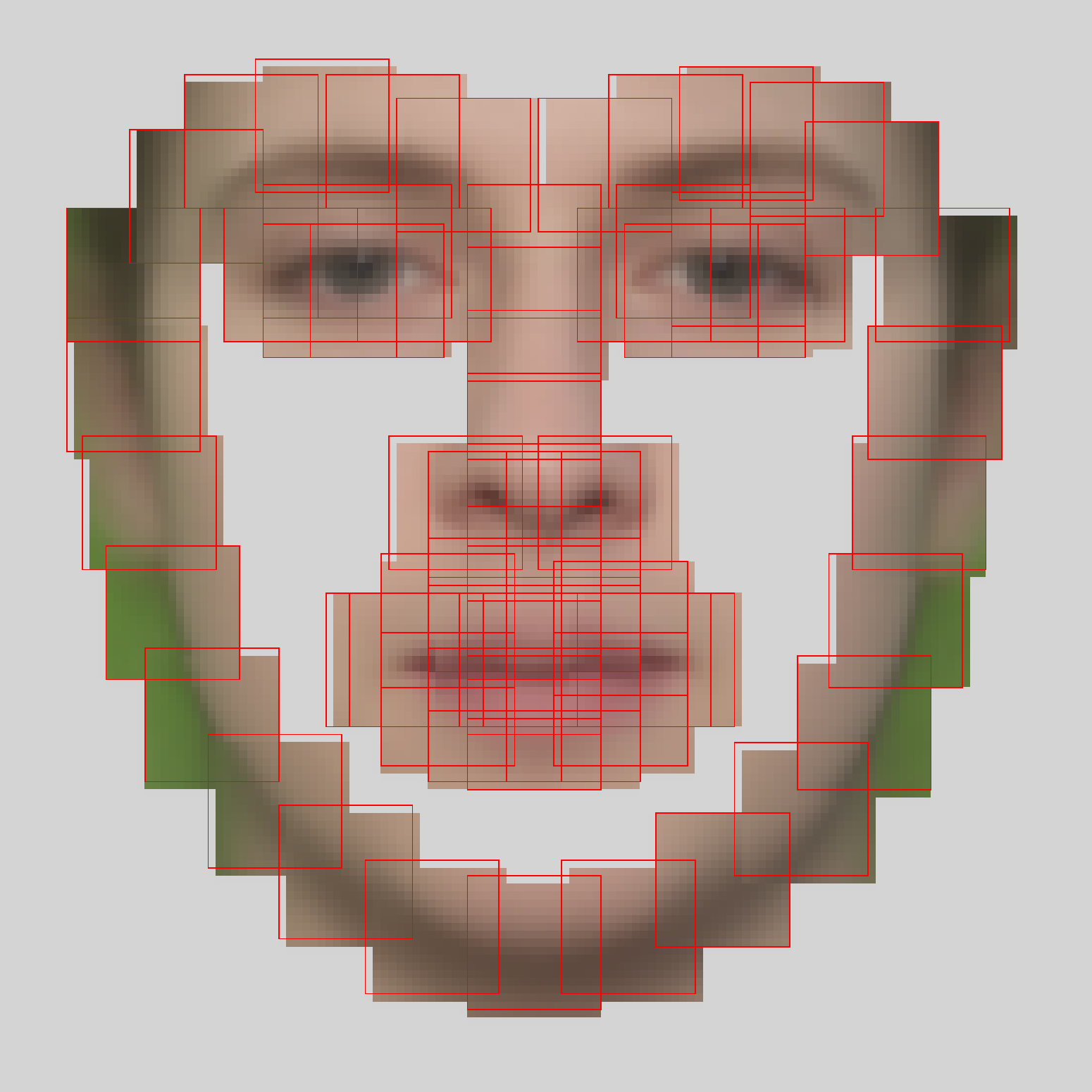}
}
\subfloat[Patch SIFT]{
	\includegraphics[width=0.5\linewidth]{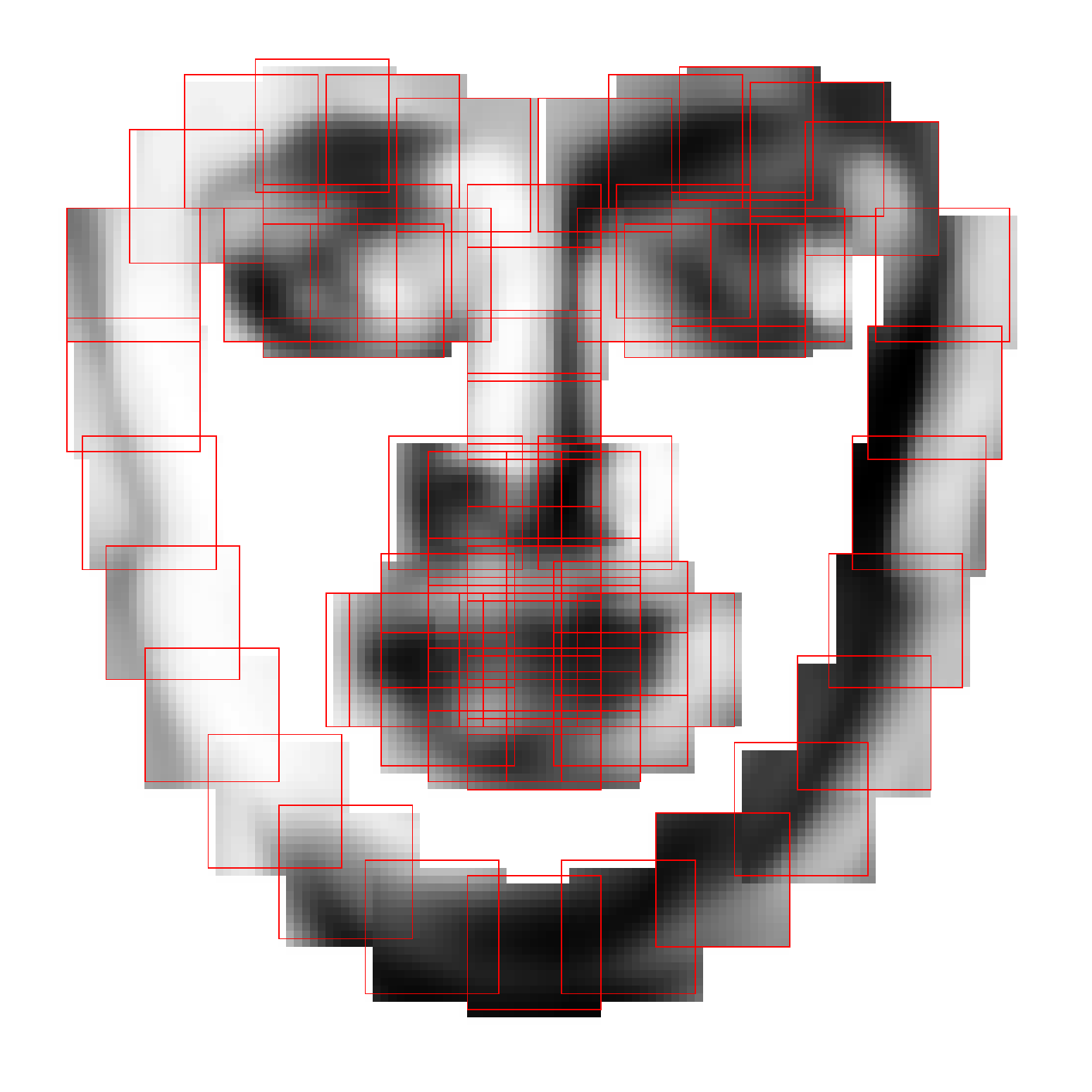}
}

\caption{Overview of AAM types by warp and feature used.\\ The \emph{Patch} models are evaluating local neighborhoods of the landmarks instead of the entire appearance. The SIFT descriptors are robust alternatives to raw pixel intensities, where no additional operation is applied (no-op).}
\label{fig:aams}
\end{figure}

Active Appearance Models (AAM), introduced in \cite{Cootes1998} and streamlined in \cite{Matthews2004}, are state-of-the-art techniques for deformable object modeling. The robustness of AAMs has greatly improved since these early publications via several factors: better fitting algorithms \cite{Alabort-i-Medina2017}, feature-based image descriptors \cite{7104116} and patch models \cite{6909635} (portrayed in Figurge~\ref{fig:aams}). These improvements are fairly recent, yet remarkable efforts have been invested to make them available in an open-source project \cite{menpo}.



As the recent AAM developments have been mostly oriented on fitting performance, the recognition performance of the AAM-based features on lipreading tasks once more becomes uncharted territory. AAMs have been applied to lipreading of simple tasks, such as isolated words \cite{6613979, 982900, 4782036} or small-vocabulary command sentences \cite{lan2009}, while the very few attempts on large-vocabulary speech are performed on the IBM ViaVoice dataset which is not publicly available \cite{review_pota}.


The main contribution of this paper is a direct comparison between AAM and Discrete Cosine Transform (DCT)-based visual features on TCD-TIMIT \cite{tcdtimit}, a publicly available audio-visual dataset aimed at large vocabulary continuous speech recognition (LVCSR). We also present an automatic procedure to train AAMs from estimates of pre-trained models, eliminating the need for manual annotations and making it applicable on any dataset. To encourage reproducibility, we make our code publicly available \footnote{http://www.mee.tcd.ie/\texttildelow sigmedia/Resources/PyVSR}.

The rest of the paper is organized as follows. In Section~\ref{sec:visfeat} we present the mathematical formulation of our visual feature processing front-ends. In Section~\ref{sec:methodology} we describe the steps taken to train AAMs and fit them to the data. Section~\ref{sec:experiments} presents our experiments, and we draw the conclusions in Section~\ref{sec:analysis}.

\section{Visual features}
\label{sec:visfeat}
\subsection{DCT}

The Discrete Cosine Transform (DCT) represents a standard choice for visual feature extraction in many lipreading tasks \cite{review_pota, review_zhou}. Although aimed at compressing the energy of a signal, it often outperformed algorithms tuned to maximize the classification accuracy, so it is used here as a baseline method.

To obtain a DCT-based feature in our framework, a region of interest (ROI) has to be first localized and isolated from the full-sized image.
As the initial work \cite{tcdtimit} provided extracted mouth ROIs, we obtained their coordinates through cross-correlation-based template matching, so we could apply different post-processing steps.
The extracted ROI is converted to grayscale, then downsampled to 36x36 pixels using cubic interpolation, and finally a 2D DCT transform is applied. The feature vector is made of the first 44 coefficients (without the DC coefficient) chosen in a zig-zag pattern and is concatenated with the first and the second derivatives. The derivatives are computed using a central finite differences scheme that is fourth order accurate, and the same order is preserved at the boundaries by using forward and backward schemes.

Since we are keeping the feature size constant, there is a trade-off between the frequency range captured by the selected DCT coefficients and the granularity of the representation. The choice for the window size was made experimentally, after trying values of 24, 28, 32, 36 and 40 pixels per side.

\subsection{AAM}

An AAM is a deformable statistical model of shape and appearance that learns the variance of an annotated set of training images. The shape consists of a set of landmarks $\mathbf{s} = [x_1,y_1,...,x_N,y_N]$ placed on the object to be modeled, which are a priori aligned using Generalized Procrustes Analysis to reduce the effect of translation, rotation and scaling. Applying Principal Component Analysis (PCA) on the set of aligned training shapes leads to a shape model expressed as:

\begin{equation}
\mathbf{s} = \bar{\mathbf{s}} + \sum_{i=1}^{n} p_i s_i = \bar{\mathbf{s}} + \mathbf{S}\mathbf{p}
\end{equation}

where any shape $s$ is a linear combination of the shape eigenvectors $s_i$ with the weights $p_i$ also known as shape parameters, plus the mean shape $\bar{s}$.

To construct the appearance model, the pixels within the training shapes are first warped to their corresponding locations in a common reference shape (typically the mean shape $\bar{s}$) using techniques such as piecewise affine warping or thin plate splines. PCA is applied again on the serialized warped image, such that any appearance $A(x)$ could be expressed as a mean appearance $\bar{A}(x)$ plus a linear combination of the appearance eigenvectors $A_i(x)$:

\begin{equation}
\mathbf{A}(x) = \bar{\mathbf{A}}(x) + \sum_{i=1}^{m}c_i A_i(x) = \bar{\mathbf{A}}(x) + \mathbf{A}\mathbf{c}
\end{equation}

where the weights $c_i$ denote the appearance parameters.

Since the number of parameters is as large as the number of landmarks and the number of pixels respectively, a trade-off can be made between the representation power of the models and the size of the parameter vectors by analyzing the cumulative ratio of the corresponding eigenvalues.

For unlabeled images, when a good initialization of the shape can be provided (e.g. the mean shape aligned on a face localized using a face detector), several fitting algorithms can be applied to iteratively update the parameters that minimize an error between the given image and the model instance. In \cite{Alabort-i-Medina2017}, such algorithms are classified with respect to the cost function, type of composition and optimization method. The parameters obtained at the last iteration constitute the foundation of the AAM-based visual features.
\setlength{\belowcaptionskip}{-10pt}
\begin{figure}[t]
  \centering
  \includegraphics[width=\linewidth]{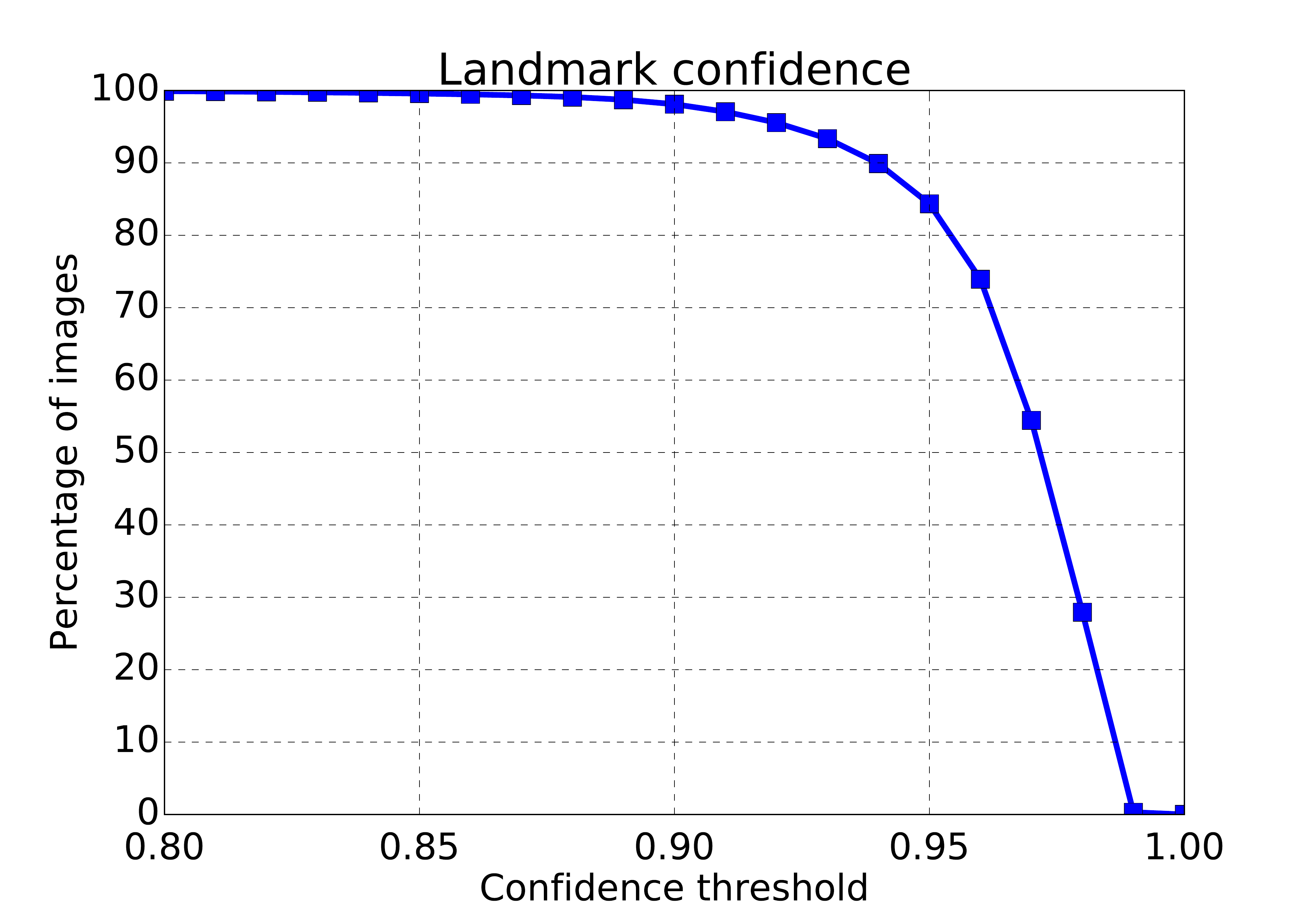}
  \caption{OpenFace landmark confidence on TCD-TIMIT}
  \label{fig:openf_conf}
\end{figure}
\setlength{\belowcaptionskip}{0pt}
\section{Methodology}
\label{sec:methodology}

\subsection{Dataset}

We have used the TCD-TIMIT dataset \cite{tcdtimit} for our experiments. The sentences of TCD-TIMIT are designed for good coverage of phoneme pairs in English, implicitly providing realistic viseme contexts, thus well suited for a large vocabulary lipreading task.

To make our results comparable to \cite{tcdtimit}, we used an identical setup for the speaker-dependent scenario. Hence, we worked on the subset of 56 speakers with Irish accents, each speaker contributing with 67 sentences for training and 31 sentences for testing. For labels, we reuse the transcription file made of sequences of 12 viseme classes, as it was based on the top performing phoneme-to-viseme mapping in the work of \cite{p2vmap}.

\subsection{AAM training}
\label{sec:aamtrain}

An annotated set of images is required to train AAMs. Previously, this has been a time-consuming step for most datasets.
In \cite{lan2009} and \cite{bear1}, a few frames per speaker are manually annotated, then person-specific AAMs are trained and fitted on the remaining frames. In addition, the final parameters are obtained by projecting the shapes and appearances onto the PCA subspace, which would roughly be equivalent to a Sum of Squared Differences (SSD) formulation of the cost function.

To eliminate the need for manual labor, we propose an automatic procedure to train our models. The open-source tool OpenFace \cite{openface} was used to get 68 facial landmark estimates for each frame, storing at the same time their confidence scores as returned by the tool. We then analyzed the cumulative distribution of these confidence scores on our dataset, shown in Fig.~\ref{fig:openf_conf}. This reveals an overall high confidence, which means that most frames have reliable labels. From a visual inspection we observed that most landmarks above a confidence score of 0.9 were very accurate, with the exception of the lips region.

Training generative models such as AAMs with a massive amount of similar data, such as consecutive video frames, leads to poor performance in practice, so we apply a sampling strategy. Taking the faces that get detected successfully and that have a high confidence score, we sort them by the amount of lip  opening (distance between the upper and lower lips). We then sample between 3 to 6\% of the sorted frames at evenly spaced intervals. For TCD-TIMIT we decided to use a confidence threshold of 0.94 to train our models, which kept 90\% of the frames. In addition, we randomly selected only 5 training sentences per speaker from the available of 67, further reducing the training data size to a total of around 1100 frames. The reference shape of the AAM was chosen as the mean shape from the first video in the dataset. We will refer to these models as \emph{global}, since they use training data from each volunteer. The models built from the training samples of a single person will be coined \emph{person-specific}.

All the previous attempts at lipreading with AAMs have used the original formulation where the entire appearance texture within the landmark area was modeled. It has been shown that learning only small patches around the landmarks leads to robust models that outperform the state-of-the-art at fitting to unseen faces \cite{6909635}. We considered both approaches, coined \emph{Holistic} and \emph{Patch} AAMs in \cite{menpo} (and illustrated in Fig.\ref{fig:aams}), in order to compare their fitting and classification performance. In addition to the traditional pixel intensities for appearance features (denoted in this work and in \cite{menpo} as \emph{no-op}), we also considered SIFT \cite{Lowe2004} image features, which were shown to largely outperform popular alternatives at fitting to unconstrained images, requiring at the same time fewer appearance components \cite{7104116}.

Modeling only a part of the face can be beneficial for lipreading \cite{7074009, 4782036}, since the PCA energy would better describe the subtle movements. Yet, as the area being modeled gets smaller, is it expected to see an increase in fitting error. We built two additional models, one for the lips area only, and another for the whole chin and mouth area (further denoted as \emph{chin}), the latter being chosen as a trade-off between relevancy to lipreading and fitting performance. The face and the chin models use a pyramid of three resolution levels (25\%, 50\%, 100\%), while the lip models only use the last two.

Other important parameters for our models were the image rescaling to a diagonal of $\approx$ 150 pixels at full scale, 40 and 150 shape and appearance components respectively, and patch sizes of 17x17 pixels around landmarks for the Patch models.

Table~\ref{tab:kept} shows how well our models were able to represent the appearance of the training data. High values of the kept variance imply that model is able to reconstruct accurately any given face, provided that the optimization algorithm finds the right parameters. More variance was kept using pixel intensities than SIFT features, as the color images have only three channels while SIFT has eight, thus more data is being modeled.
The variance kept by the shape eigenvectors was close to 100\% using 40 components, suggesting that there are strong correlations between the landmark locations.

\begin{table}[th]
\centering
\caption{Percentage of kept variance for the appearance models using 150 appearance components}
\label{tab:kept}
\begin{tabular}{|l|l|l|l|l|l|}
\hline
\multirow{2}{*}{\textbf{\begin{tabular}[c]{@{}l@{}}Model $\rightarrow$\\$\downarrow$ Part \end{tabular}}} & \multicolumn{2}{c|}{\textbf{Holistic}} & \multicolumn{2}{c|}{\textbf{Patch}} & \multirow{2}{*}{\textbf{Scale}} \\ \cline{2-5}
                                                                                     & \textbf{no-op}     & \textbf{SIFT}     & \textbf{no-op}    & \textbf{SIFT}   &                                 \\ \hline
\multirow{3}{*}{\textbf{face}}                                                       & 96.6              & 78.7               & 83.1             & 63.0             & \textbf{25\%}                   \\ \cline{2-6} 
                                                                                     & 96.8              & 79.2               & 87.6             & 71.1             & \textbf{50\%}                   \\ \cline{2-6} 
                                                                                     & 93.2              & 76.9               & 82.8             & 74.7             & \textbf{100\%}                  \\ \hline \hline
\multirow{3}{*}{\textbf{chin}}                                                       & 97.9              & 75.9               & 82.8             & 56.4             & \textbf{25\%}                   \\ \cline{2-6} 
                                                                                     & 97.1              & 73.4               & 87.4             & 65.0             & \textbf{50\%}                   \\ \cline{2-6} 
                                                                                     & 93.6              & 70.1               & 83.9             & 69.6             & \textbf{100\%}                  \\ \hline \hline
\multirow{2}{*}{\textbf{lips}}                                                       & 95.4              & 72.2               & 89.2             & 61.6             & \textbf{50\%}                   \\ \cline{2-6} 
                                                                                     & 91.6              & 68.5               & 90.9             & 65.9             & \textbf{100\%}                  \\ \hline
\end{tabular}
\end{table}

\subsection{AAM feature selection}
\label{sec:aamfeat}

The AAM fitting process consists in the optimization of a cost function (typically the error between a given image and the AAM reconstruction) with respect to the shape and appearance parameters, provided that a good initialization is available. The shape was initialized using the \emph{dlib} face detector implemented in \emph{menpo} \cite{menpo} by aligning the mean shape with the face bounding box. The Wiberg Inverse Compositional (WIC) algorithm was chosen for the optimization problem, as it was shown to be an efficient alternative to state of the art algorithms \cite{Alabort-i-Medina2017}. We ran 10 iterations of WIC for the first two resolution scales and 5 more for the full resolution model, with an important exception of the Holistic no-op model that needed 20 iterations at the lowest scale in order to converge more often. For the \emph{chin} and \emph{lips} models, the shapes were initialized from a subset of the final \emph{face} shape, iterating 10 more times per resolution scale to make room for corrections.

We considered the shape and the appearance parameters after the last iteration as feature vectors, either taken separately or concatenated, and we also considered the first derivative of the appearance alone or the concatenation. Among these five features, the highest performance was achieved by the latter, which was our default choice in the subsequent experiments. The first four shape parameters were discarded, as they represented the global similarity transform used for normalization.

It is worth mentioning that fitting is a slow process, taking almost one day to process the files of a single speaker using the four face models alone in \emph{menpo}. We ran the fitting process on a HPC cluster made of 16 nodes and 40 cores, achieving a theoretical speedup factor of 160.

\subsection{Viseme recognition}

Our monoviseme recognizer was implemented in HTK 3.5 \cite{htkbook}, following the procedure described in \cite{tcdtimit} as close as possible.
For each of the 12 viseme classes we have built 3-state left-to-right Hidden Markov Models (HMMs) with mixtures of 20 diagonal covariance Gaussian densities per state, initialized in flat start mode with \emph{HCompV}. Additionally, for the silence state viseme  we have added backward and skip transitions. Finally, we have applied 5 runs of embedded training using \emph{HERest} for every increment of the mixture components.

The reported correctness and accuracy results are computed using \emph{HResult} between the ground-truth transcriptions provided with TCD-TIMIT and the predicted ones.

No language model was used. This allows a comparison of the raw lipreading ability of the feature sets.

\section{Experiments}
\label{sec:experiments}

\subsection{Fitting performance}
\label{sec:expfitt}

The overall system performance relies first on the accuracy of landmark localization on unseen faces. In this experiment we compare the performance of our face AAMs in terms of face-normalized point-to-point Euclidean error between the WIC fitter prediction and the ground-truth shapes.

Although the ground-truth labels are not perfect, having a high confidence rate as in Figure \ref{fig:openf_conf} leaves little room for noise. We obtained almost identical results when considered the fitting performance only on the frames above 0.94 confidence.

\begin{figure}[t]
  \centering
  \includegraphics[width=\linewidth]{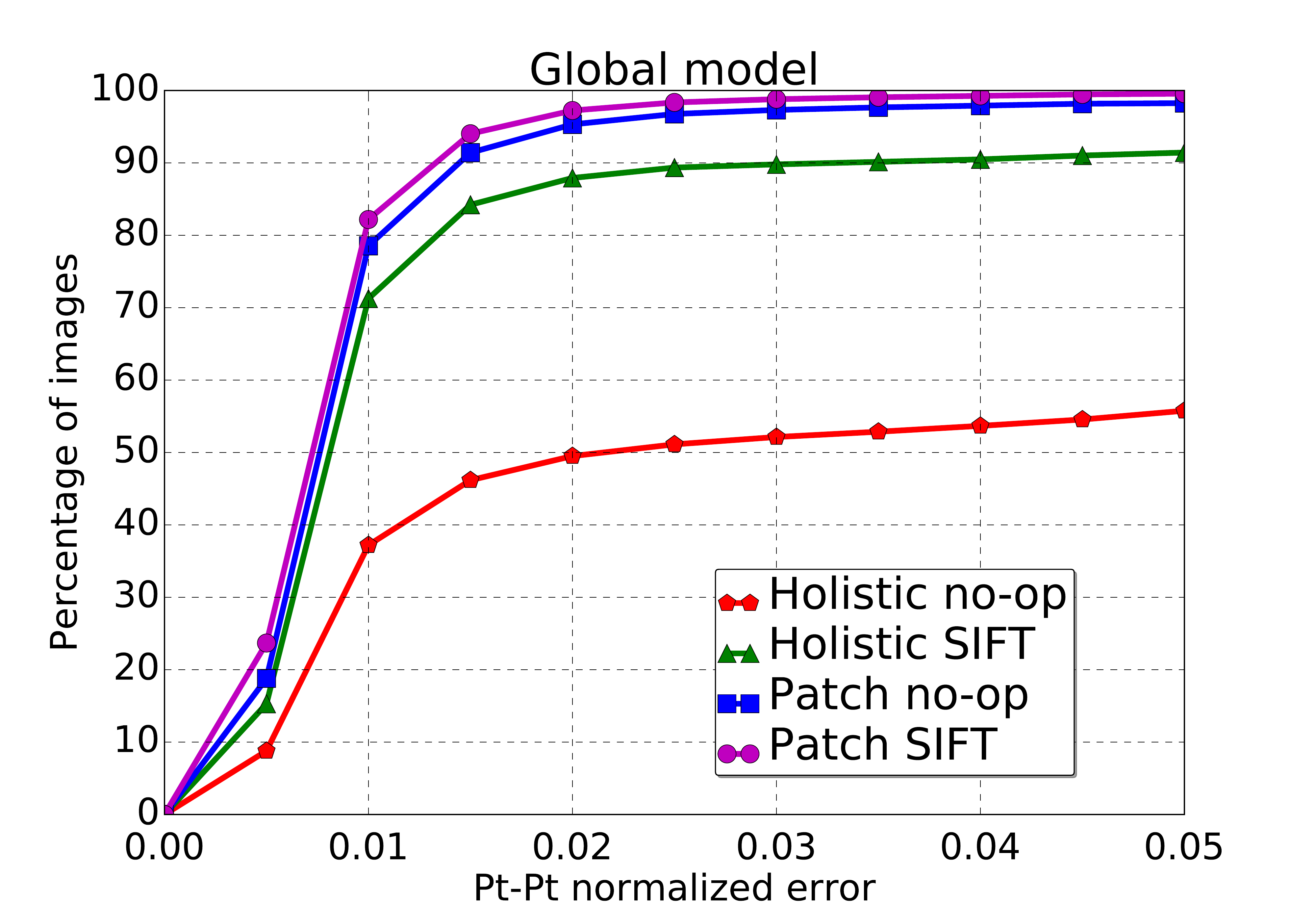}
  \caption{AAM fitting convergence using global face models (trained on the full set of volunteers)}
  \label{fig:cvg_all}
\end{figure}

\begin{figure}[t]
  \centering
  \includegraphics[width=\linewidth]{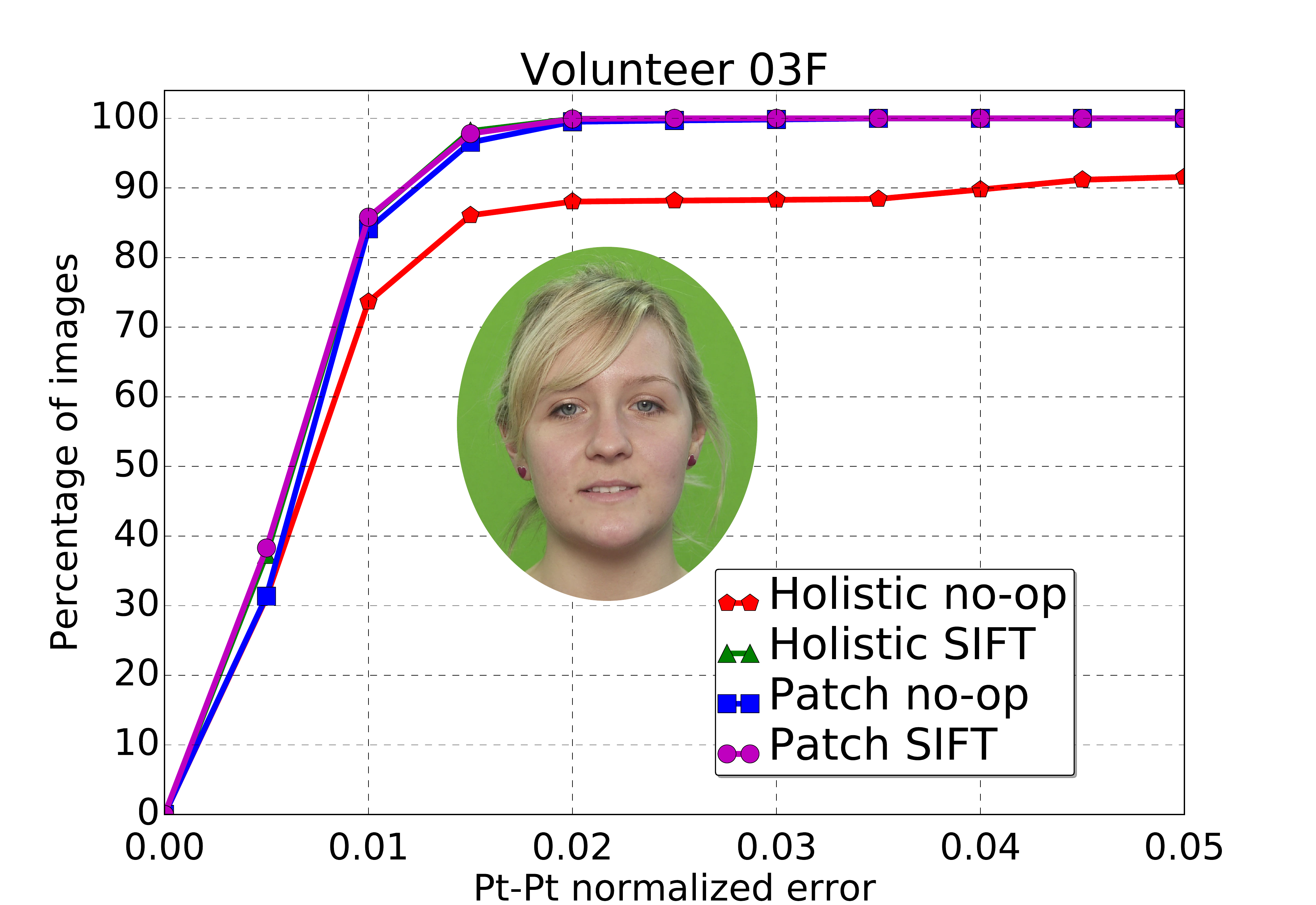}
  \caption{AAM fitting convergence using a person-specific model for volunteer 03F}
  \label{fig:cvg_03f}
\end{figure}

\begin{figure}[t]
  \centering
  \includegraphics[width=\linewidth]{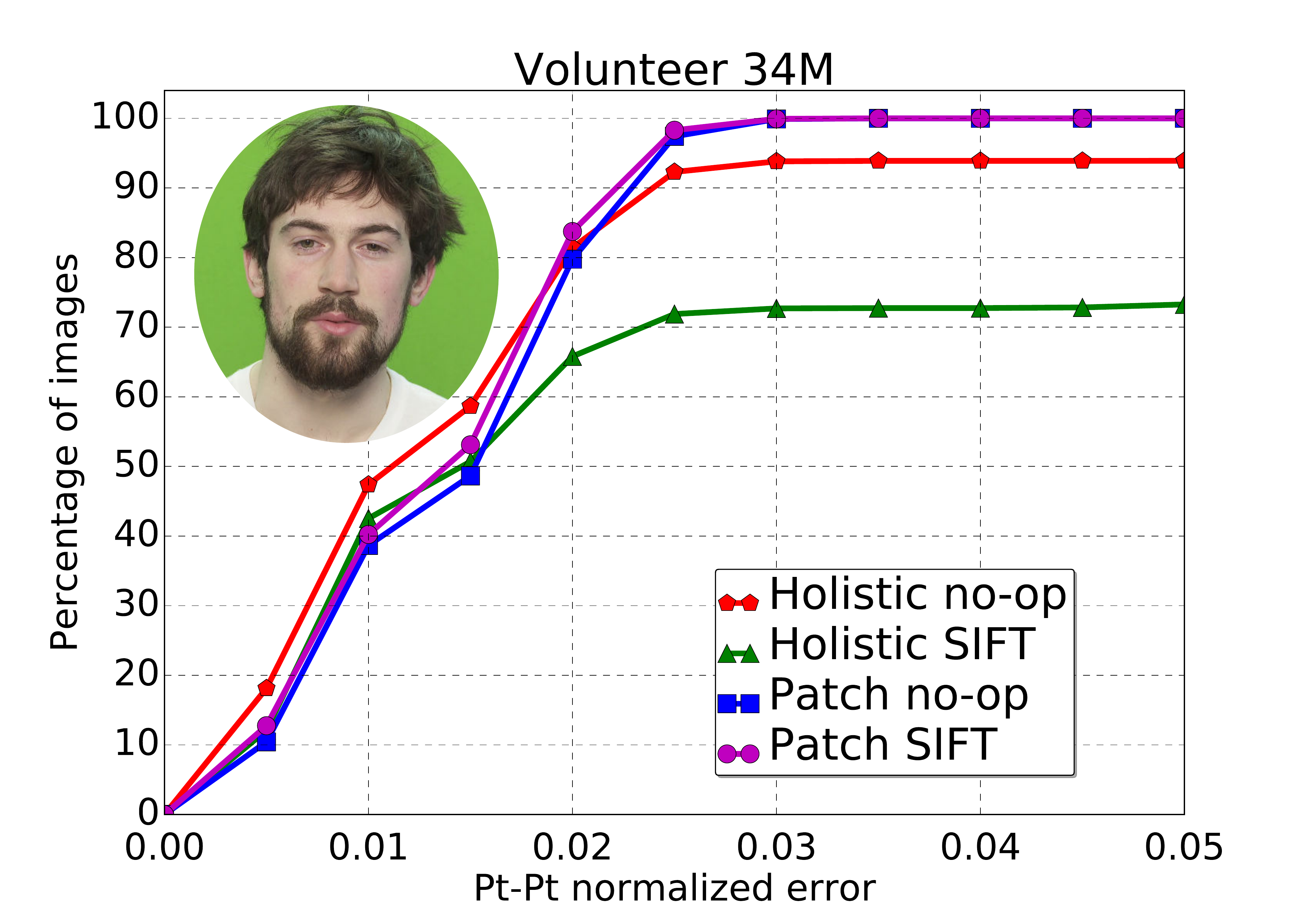}
  \caption{AAM fitting convergence using a person-specific model for volunteer 34M}
  \label{fig:cvg_34m}
\end{figure}

Figure~\ref{fig:cvg_all} shows the proportion of frames fitted with an error lower than a certain value, using the global face models, while Figures~\ref{fig:cvg_03f}-\ref{fig:cvg_34m} show the same information using person-specific AAMs of two volunteers.
The two speakers modeled individually were drawn from the top/bottom 10 performers in \cite{tcdtimit}, where volunteer \emph{03F} was considered easier to lipread than \emph{34M}, which had a full beard and moustache.
 
The Holistic models were outperformed by the Patch models in almost all cases, with the exception of volunteer \emph{03F} where \emph{Holistic SIFT} managed to match them, although for volunteer \emph{34M} it couldn't cope well with the facial hair. Both Patch models achieved a convergence rate above 95\% for an error of 0.02 and were almost indistinguishable in performance, demonstrating their robustness not only for fitting to unseen frames, but also when trained from less perfect landmarks.

\setlength{\belowcaptionskip}{-8pt}
\setlength{\belowcaptionskip}{0pt}

\setlength{\belowcaptionskip}{-8pt}
\setlength{\belowcaptionskip}{0pt}
\setlength{\belowcaptionskip}{-8pt}
\setlength{\belowcaptionskip}{0pt}
\setlength{\belowcaptionskip}{-8pt}
\begin{figure}
  \centering
  \includegraphics[width=\linewidth]{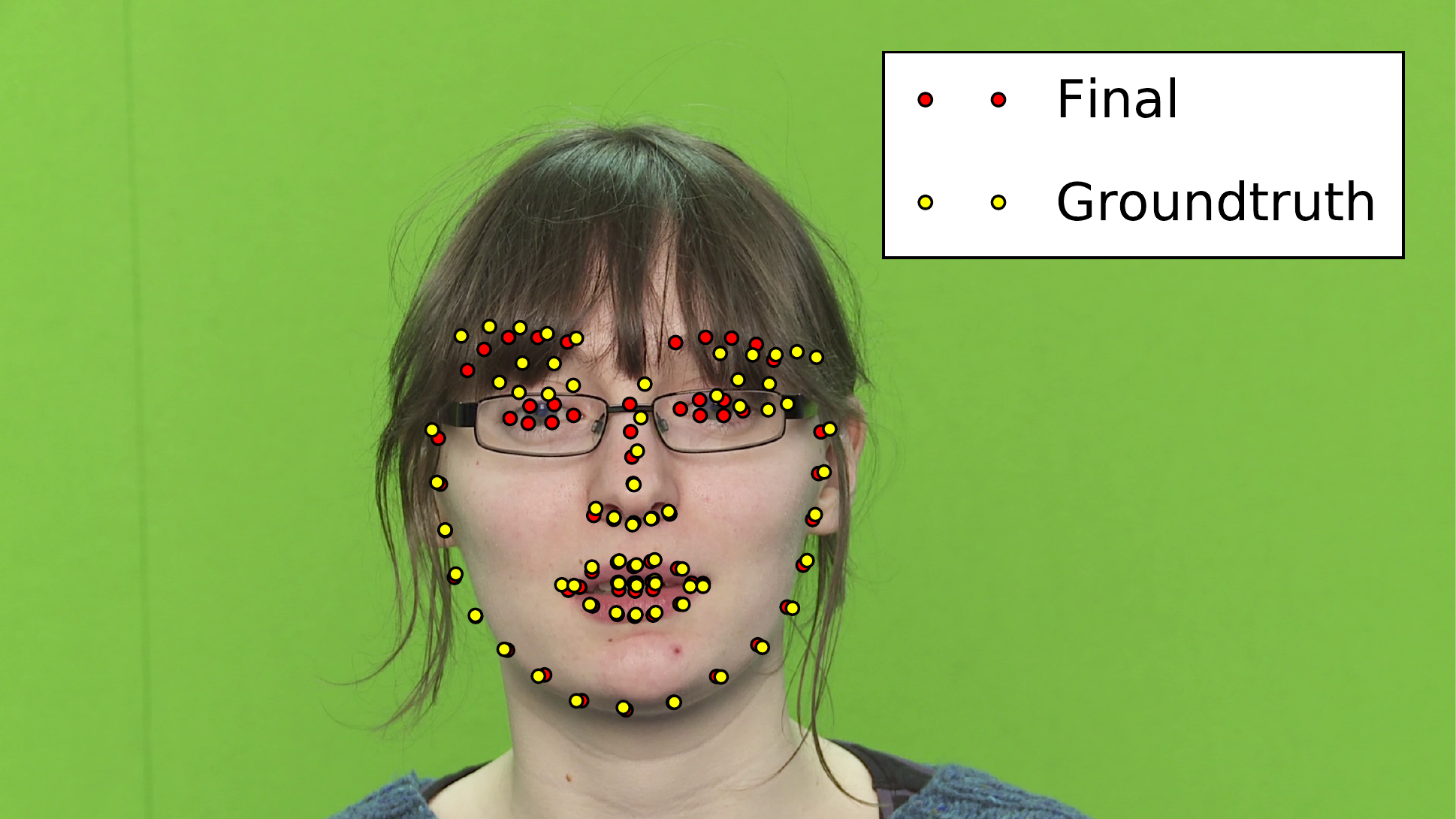}
  \caption{{Landmark correction for volunteer 05F wearing glasses and with the eyebrows occluded}}
  \label{fig:alva}
\end{figure}
\setlength{\belowcaptionskip}{0pt}

In most cases, AAMs were able to improve the pre-trained OpenFace estimates where the confidence score was low. 
One such example is shown in Figure \ref{fig:alva}, where the eyes and the eyebrows landmarks were corrected for volunteer \emph{05F} wearing glasses with the eyebrows not visible. This leads to a better parametrization of the fitter for faces that are otherwise more challenging to landmark.

\subsection{Recognition performance}

We now focus on the recognition results obtained by training HMMs in a speaker-dependent scenario, thus using 67 training sentences from each volunteer and testing on their remaining 31 unseen sentences. The predicted viseme sequence is computed using the HTK tool \emph{HVite}. 

\begin{figure}[t]
  \centering
  \includegraphics[width=\linewidth]{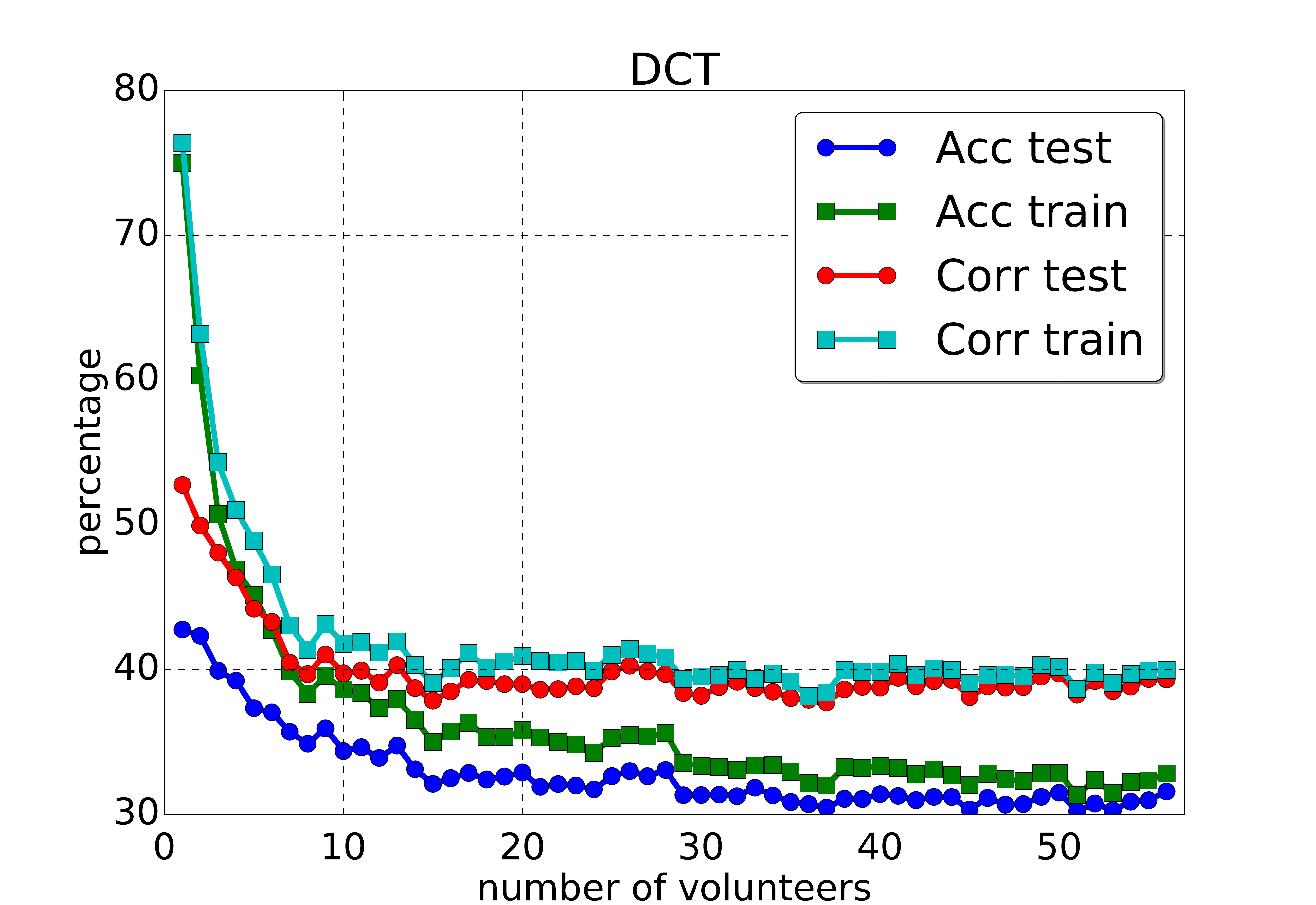}
  \caption{Correction and Accuracy scores for DCT features}
  \label{fig:dct}
\end{figure}

In Figure~\ref{fig:dct} we plot the correctness and accuracy scores returned by \emph{HResults} for an increasing number of volunteers added to the system (ordered by their alphanumeric IDs). The accuracy on the entire set of volunteers (31.59\%) is 3\% below the one obtained in \cite{tcdtimit}. An increase of 1-2\% was possible when we interpolated the features to double the rate and used 4-state HMMs, but we reverted to the original settings to have a fair comparison with the AAM features.

\setlength{\belowcaptionskip}{-8pt}

\setlength{\belowcaptionskip}{0pt}

\begin{figure}[t]
  \centering
  \includegraphics[width=\linewidth]{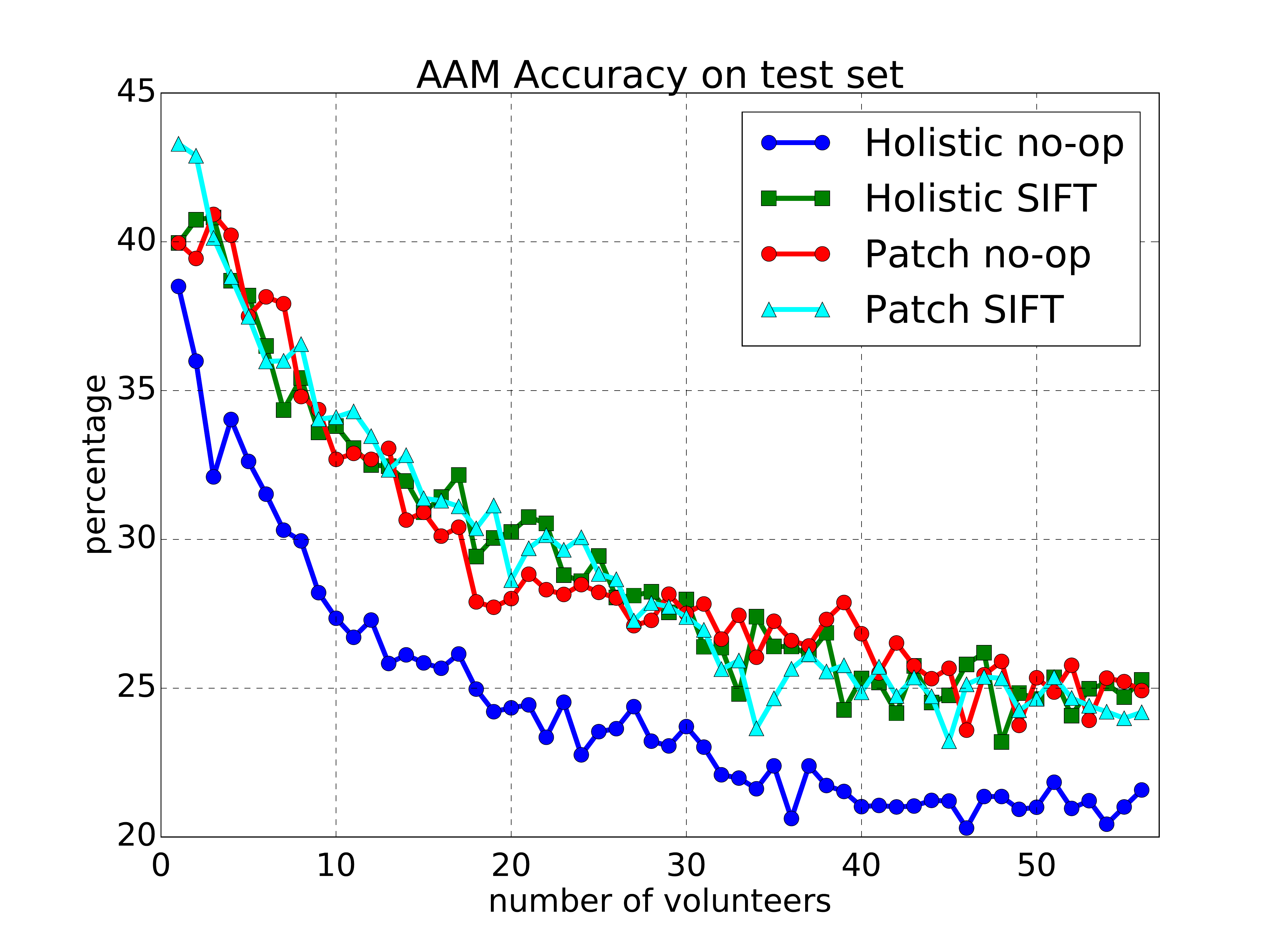}
  \caption{Accuracy scores for AAM features}
  \label{fig:aam4}
\end{figure}

In Figure~\ref{fig:aam4} we show the accuracy obtained using AAM-based features and an identical HMM recognition framework. As anticipated, the \emph{Holistic no-op} model has the lowest accuracy, since less than 60\% frames converged on average. The other three models perform similarly, yet reaching an accuracy of $\approx$ 25\% on the entire set, significantly lower than DCT. 

\setlength{\belowcaptionskip}{-8pt}
\setlength{\belowcaptionskip}{0pt}

\begin{figure}[t]
  \centering
  \includegraphics[width=\linewidth]{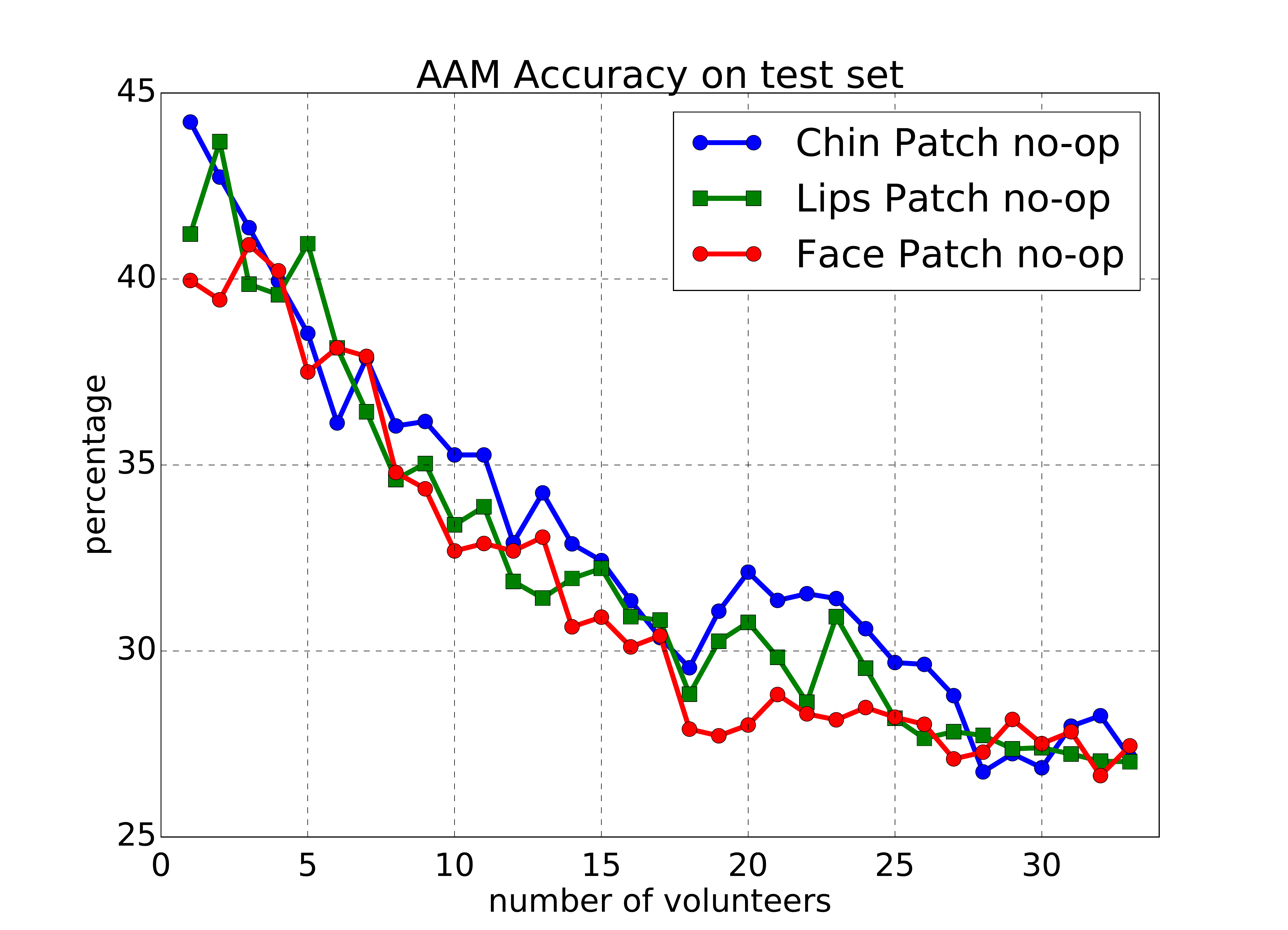}
  \caption{Performance of the \emph{chin} and \emph{lips} models}
  \label{fig:chinlips}
\end{figure}

We repeated the experiment with features extracted using the two part models, \emph{chin} and \emph{lips}, on a subset of the first 33 volunteers, following the process described in Section \ref{sec:aamfeat}. The results are displayed in Figure~\ref{fig:chinlips}, showing the \emph{chin} model to perform only marginally better, although the decreasing trend remains. This small increase comes with the cost of doubling the processing time, as it requires a cascade of two fittings.

\setlength{\belowcaptionskip}{-8pt}
\setlength{\belowcaptionskip}{0pt}

\subsection{Speaker-specific models}

In order to see how much the quality of the AAM impacts the viseme recognition accuracy, we tested the case of person-specific AAMs for the two volunteers described in Section~\ref{sec:expfitt}. If there was a problem with the global model, we should notice a significant increase in accuracy when switching to person-specific models.
Table~\ref{tab:spec} shows the viseme recognition results obtained with both specific and global models for these two speakers, along with the DCT baseline. We could not find a significant advantage of the person-specific models, hence at this stage it would not be useful to attempt adapting a global AAM to particular faces in order to gain a performance boost.

\begin{table}[]
\centering
\caption{Recognition performance for person-specific models versus the global models. Since there was less data available for individual speakers, the highest values were obtained on average with 14 Gaussian mixture densities}
\label{tab:spec}
\begin{tabular}{|c|l|l|l|l|}
\hline
\multicolumn{1}{|l|}{\multirow{2}{*}{\textbf{\begin{tabular}[c]{@{}l@{}}Speaker $\rightarrow$\\$\downarrow$ Part \end{tabular}}}} & \multicolumn{2}{c|}{\textbf{03F}}                                      & \multicolumn{2}{c|}{\textbf{34M}}                                      \\ \cline{2-5} 
\multicolumn{1}{|l|}{}                                                                                     & \multicolumn{1}{c|}{\textbf{Corr}} & \multicolumn{1}{c|}{\textbf{Acc}} & \multicolumn{1}{c|}{\textbf{Corr}} & \multicolumn{1}{c|}{\textbf{Acc}} \\ \hline
\multicolumn{5}{|c|}{Specific AAM}                                                                                                                                                                                                                           \\ \hline
\textbf{face}                                                                                              & 51.84                              & 42.62                             & 51.63                              & 43.24                             \\ \hline
\textbf{chin}                                                                                              & 52.62                              & 45.24                             & 53.11                              & 40.18                             \\ \hline
\textbf{lips}                                                                                              & 50.87                              & 43.98                             & 52.62                              & 43.14                             \\ \hline
\multicolumn{5}{|c|}{Global AAM}                                                                                                                                                                                                                             \\ \hline
\textbf{face}                                                                                              & 53.11                              & 44.66                             & 51.04                              & 41.76                             \\ \hline
\textbf{chin}                                                                                              & 53.88                              & 43.50                             & 51.53                              & 41.95                             \\ \hline
\textbf{lips}                                                                                              & 52.43                              & 44.27                             & 53.70                              & 42.15                             \\ \hline
\multicolumn{5}{|c|}{}\\ \hline
\textbf{DCT}                                                                                              & 54.66                              & 46.80                             & 47.88                              & 39.68                             \\ \hline
\end{tabular}
\end{table}

\section{Discussion}
\label{sec:analysis}

In this paper we have explored the performance of hand-crafted visual features for a LVCSR lipreading task in a traditional HMM framework. We first computed DCT-based features for a baseline, reaching a similar result as in \cite{tcdtimit}. Then we trained several AAMs using an automatic procedure and fitted them to each video frame to obtain the AAM-based features.

A first finding is that AAM features do not outperform the DCT ones in an identical recognition framework. This has been reported before on IBM ViaVoice \cite{review_pota}. This dataset has 290 subjects and over 50 hours of speech. However their approach was to rescore audio-only lattices with visual unit HMMs. Their scenario therefore bypassed the issue of using visual features to find the viseme boundaries. On the other hand, the study of \cite{lan2009} found AAM better than DCT on a lipreading task with a small vocabulary of 51 words, where word-level HMMs were used. Later work from the same authors in \cite{lan2010} reported results on a corpus of 12 speakers, each speaking 200 sentences from a vocabulary totalling 1000 words. Again AAM outperformed DCT, but the approach made use of Linear Discriminant Analysis requiring frame-aligned viseme labels, while the facial landmarks were obtained semi-automatically from person-specific trackers. Another study \cite{6613979} used speaker-specific normalization that makes the results less comparable. This is the most comprehensive comparison between DCT and state-of-the-art AAM that we are aware of.

The reported results are obtained using a visual speech model only, allowing raw performance comparison of the extracted features. Adding a simple bigram language model improves the viseme recognition accuracy by up to 10\% for the AAM features, and 3\% for DCT, narrowing the performance gap to less than 1\%. 

Both AAM and DCT perform a basis decomposition of the image, although the first considers the eigenvectors specific to a training set of images, while the latter uses a fixed frequency decomposition. Since both transforms are not optimized for classification, e.g. maximizing the separability between classes, this suggests that the raw parameters are not necessarily ideal features, requiring further processing to find person-independent cues. This is reinforced by the fact that \emph{Patch AAMs} obtained a high convergence rate at fitting to unseen images, so the parameters should contain meaningful information.

Modeling a subset of the face has only shown minor improvements of the recognition accuracy. The \emph{chin} model seems to have a slightly better advantage versus the \emph{lips} one, and this could be explained by two factors. The extra iterations of the part model ensured a more accurate fitting where there were more control points available. Also, the chin area contains additional visemic information, as speech articulators are not limited to the lips region. 

In this context, we could also question the suitability of HMMs for LVCSR lipreading. We plan to reproduce our experiments on simpler datasets such as GRID \cite{gridset}, CUAVE \cite{cuaveset}, and also on OuluVS2 \cite{ouluvs2} which is similar to TCD-TIMIT. A thorough analysis of the HMM suitability would require debugging down to the Baum-Welch and Viterbi algorithms to understand the fail cases. Another informative experiment would be to replace the HMM framework for pattern recognition with a Long short-term Memory (LSTM) one, while reusing exactly the same features, as it could reveal a bottleneck at the recognition level and not the feature one.

An important conclusion about AAMs is that the \emph{Patch} models, especially when combined with SIFT image descriptors, are able to achieve a much higher fitting and implicitly recognition accuracy than the traditional \emph{Holistic} ones that have been used so far in lipreading. As shown in \cite{6909635}, their robustness is conspicuous when trained on unconstrained in-the-wild faces, making them more suitable candidates for realistic lipreading scenarios.

\section{Acknowledgements}
The ADAPT Centre for Digital Content Technology is funded under the SFI Research Centres Programme (Grant 13/RC/2106) and is co-funded under the European Regional Development Fund. This work was also supported by TCHPC.
\bibliographystyle{IEEEtran}

\bibliography{mybib.bib}

\end{document}